\documentclass[sort&compress,10pt]{iopart}

\usepackage{graphicx}
\usepackage{iopams} 
\usepackage{siunitx}
\usepackage[T1]{fontenc} %
\usepackage[english]{babel}
\usepackage{bm}

\usepackage[usenames,dvipsnames]{color}
\usepackage{cite}
\usepackage{graphicx}
\usepackage{hyperref}
\usepackage[utf8]{inputenx}
\usepackage{xspace}
\usepackage{siunitx}

\usepackage[scaled=0.9]{helvet}      %

\usepackage{mathptmx}    %

\newcommand\potfit{\textit{potfit}\xspace}

\begin{document}

\title[\sffamily Uncertainty quantification for classical effective potentials: an extension to \potfit]{Uncertainty quantification for classical effective potentials: an extension to \potfit}

\author{\sffamily Sarah Longbottom and Peter Brommer}

\address{Warwick Centre for Predictive Modelling, School of Engineering and Centre for Scientific Computing, University of Warwick, Coventry CV4 7AL, UK}
\ead{\rmfamily \href{mailto:p.brommer@warwick.ac.uk}{\rmfamily p.brommer@warwick.ac.uk}}
\vspace{10pt}

\begin{abstract}
Effective potentials are an essential ingredient of classical molecular dynamics (MD) simulations. Little is understood of the consequences of representing the complex energy landscape of an atomic configuration by an effective potential or force field containing considerably fewer parameters. The probabilistic potential ensemble method has been implemented in the \potfit force matching code. This introduces uncertainty quantification into the interatomic potential generation process. Uncertainties in the effective potential are propagated through MD to obtain uncertainties in quantities of interest, which are a measure of the confidence in the model predictions. 

We demonstrate the technique using three potentials for nickel: two simple pair potentials, Lennard-Jones and Morse, and a local density dependent embedded atom method (EAM) potential. A potential ensemble fit to density functional theory (DFT) reference data is constructed for each potential to calculate the uncertainties in lattice constants, elastic constants and thermal expansion. We quantitatively illustrate the cases of poor model selection and fit, highlighted by the uncertainties in the quantities calculated. This shows that our method can capture the effects of the error incurred in quantities of interest resulting from the potential generation process without resorting to comparison with experiment or DFT, which is an essential part to assess the predictive power of MD simulations.
\end{abstract}

\section{Introduction}
Our understanding of the physics underlying material properties relies on verification from computational models of materials and molecules. Such materials simulations also allow us to predict new properties and structures which can then be reproduced experimentally. In order to facilitate the modelling, interatomic potentials have long been used to circumvent the limitations in timescale and simulation size of costly first principles calculations by specifying the energy dependence on the atomic positions. This functional representation is a key constituent of molecular dynamics simulations, where the quality of the output relies predominantly on the potential employed. Currently, the systematic error incurred in using an interatomic potential is generally unknown, as is the resulting effect on quantities of interest it is used to predict, therein forming the motivation for this work.

The general idea behind interatomic potentials is that the energy of a collection of atoms can be represented by an explicit functional form or model, dependent on the atomic separation. These analytic potential functions encode the physics into the system and contain a limited number of additional parameters which are adjusted to reproduce desired quantities. The first potentials used intuitive functional forms, fitted to experimental data; however new potentials are frequently fitted to \emph{ab-initio} data such as atomic forces, energies and stresses. We use the \potfit open source force matching code to fit interatomic potentials to density functional theory (DFT) data, and subsequently quantify the uncertainty incurred in simulations using the potential, in lieu of a first principles approach.

There has been a significant recent effort in the literature to quantify uncertainty in this multiscale modelling process, with Bayesian frameworks proposed for a variety of interatomic models and force fields \cite{PhysRevLett.93.165501,UQ_MD_FF, ReaxFF}. There has also been work toward the quantification of uncertainty due to the potential fitting reference set \cite{Dallas}, as well as a proposed framework to efficiently propagate parameter uncertainties to molecular dynamics (MD) outputs \cite{UQ_framework}. More specifically, quantification of parameter uncertainty for single potentials has been undertaken in a handful of cases \cite{Silica-LJ-UQ, VOHRA2018297,LJ_alkanes, high_pressure_FF}. We add to the growing body of uncertainty quantification (UQ) work in potential development and application by providing an open source implementation of the framework in \cite{PhysRevLett.93.165501} for use in future potential development projects.

We have implemented a new module in \potfit which adds a potential ensemble functionality to the potential fitting work flow. 
For potentials fitted with \potfit, the corresponding ensemble informs of the parameter uncertainties, given the associated reference set. 

 \Sref{sec:force_matching} introduces the \potfit code, followed by an outline of the potential ensemble method in \sref{sec:pot_ensemble}. Implementation specific details are outlined in \sref{sec:implementation}. We then demonstrate in \sref{sec:application} how an ensemble can be propagated through molecular dynamics simulations, illustrating the incurred uncertainties in both equilibrium and non-equilibrium quantities of interest (QoI): the equilibrium lattice constant, elastic constants $C_{11}$, $C_{12}$ and $C_{44}$, and thermal expansion coefficient at \SI{300}{\kelvin}.

\section{Methodology}
\subsection{The \potfit force matching code}
\label{sec:force_matching}

The \potfit code \cite{potfit_url,doi:10.1080/14786430500333349,Brommer:2007:295,Brommer:2015:74002,0965-0393-25-1-014001} is an open source package implementing the force matching method \cite{Ercolessi:1994:583}, where the parameters of an interatomic potential are adjusted to optimally reproduce forces, energies and stresses typically obtained from DFT calculations. The potential parameters $\btheta=\{\theta_1,..., \theta_N\}$ either belong to an analytic potential, or are the values of the potential function at sampling points for tabulated potentials. 

In the force matching method, the deviation from the reference data is quantified by the $N$-dimensional cost function or kernel, $C(\btheta)$, with
\begin{equation}
\label{eq:cost_function}
	 C(\btheta) = \sum_{k=1}^{\rm M} a_k \left( F_k(\btheta)-F_k^0 \right) ^2 + \sum_{r=1}^{\rm N_{\rm C}}  b_r \left( A_r(\btheta) - A_r^0 \right)^2.
\end{equation}
The first sum runs over all M force components of the reference configurations in the training set, where $F_k^0$ is the set of individual atomic forces, with weighting $a_k$. Here $F_k(\btheta)$ represents the corresponding set of forces from the potential for each atom in the configuration. The fit can be enhanced with additional information about the target system energies and stresses (and optionally other quantities), represented by $A_r^0$. The quantities can be obtained through first principles calculations and can be given weights $b_r$, depending on the importance of accuracy in their descriptions given by the potential, $A_r(\btheta)$.  The best fit potential parameters $\btheta^*$ produce the minimum cost value, $C( \btheta^*)$.

The fitting procedure, by default, uses a combination of simulated annealing followed by a gradient descent method to minimise the cost function, although there are a variety of space searching algorithms implemented. Recently, \potfit was also updated to work within the OpenKIM framework \cite{OpenKIM, Tadmor2011, Tadmor2013}, providing users with easy access to a repository of fitted and tested potential models to utilise and expand.

\subsection{The potential ensemble method}
\label{sec:pot_ensemble}

In the cost landscape defined by the potential fitting procedure, there typically is significant covariance between potential parameters, hence the eigen-directions are used to define the basin curvature. The relative degrees of curvature are given by the eigenvalues of the Hessian at the cost minimum
\begin{equation}
\label{eq:hessian}
H_{ij}(\btheta^*)=\frac{\delta^{2}C(\btheta)}{\delta\theta_{i}\delta\theta_{j}}\bigg|_{\btheta=\btheta^*},
\end{equation}
where $\btheta=\{\theta_i\}_{i=1}^{\rm N}$  represents a set of interatomic potential parameters. The Hessian calculation is defined in terms of percentage change to each parameter to overcome the issue of curvature across different length scales.
Using the information about the curvature of the minimum, we can generate an ensemble of candidate potentials of varying suitability. This ensemble inherently describes the robustness of each parameter fit to the reference data, and hence their uncertainty. 

The eigenvalues of the minimal cost space Hessian often have a large spread in their magnitudes, representing a best-fit basin with vastly differing steepness along eigendirections. The hallmark of a sloppy model is that the the basin encapsulating the minimum has significantly differing degrees of curvature along the principal axes, therefore the majority of interatomic potentials fitted in \potfit fall into the category of sloppy models. Investigation into the sampling of such sloppy models has been extensively undertaken by Brown and Sethna \cite{Brown2003}, with a focus on their occurrence in systems biology. The approach has since been outlined for interatomic potentials \cite{PhysRevLett.93.165501} and forms the basis for the implementation of the uncertainty quantification in \potfit. The approach relies on generating a potential ensemble to represent the uncertainty, by drawing samples scaled in parameter directions using curvature information from the Hessian in (\ref{eq:hessian}).

Markov chain Monte Carlo (MCMC) is used to draw samples from the minimum basin. Candidate steps are generated using random displacements, taking into account information about the curvature from the eigenvalues of the minimum Hessian. In this way, larger steps are taken in sloppy directions (i.\ e.\ those associated with small eigenvalues), with smaller steps proposed in stiffer directions. 

Steps are taken in cost space, starting from the best fit parameter set, $\btheta^*$, by proposing a simultaneous perturbation to each parameter of the form

\begin{equation}
\label{eq:rvalue}
\Delta\theta_{i}=\sum_{j=1}^{\rm N}\sqrt{\frac{\rm R}{{\rm{max}} (\lambda_{j},1)}}V_{ij}r_{j} 
\end{equation}
where $\Delta\theta_i$ is the proposed change to each potential parameter $i$, R is a system dependent scaling factor and $V_{ij}$ the matrix of normalized eigenvectors of the Hessian. The parameter $\lambda_j$ is $j$-th Hessian eigenvalue and $r_j$ a normally distributed random number. 

The acceptance criteria for a MCMC step is set by a temperature, $T$, where the cost minimum is analogous to sampling at a temperature of $T=0$. The sampling temperature is by default chosen to be the ``natural'' temperature $T_0=2C(\btheta^*)/{\rm N}$ as each mode in a harmonic model contributes an energy of $T/2$ \cite{PhysRevLett.93.165501}. 

The acceptance probability of each the Monte Carlo move is calculated as 
\begin{equation}
P_{\rm acc}(\btheta_{i+1})=\cases{1&if $C(\btheta_{i+1})< C(\btheta_i)$\\
	 \rme^{-\frac{1}{T_0}(C(\btheta_{i+1})-C(\btheta_i))}&otherwise.\\}
\end{equation}
This ensures downhill moves are always accepted, and that MC moves to higher cost potentials are accepted with a probability decreasing exponentially with the increase in cost between potential parameter sets.

\section{Extensions to \potfit: uncertainty quantification}
\label{sec:implementation}

The new functionality in \potfit allows for the quantification of uncertainty in any potential fitted using the program, as well as any quantities predicted by it. The uncertainty quantification extension produces a set of accepted MCMC potentials with their attributed cost and weighting. A decorrelated subset forms an ensemble of potentials which can be used to quantify the uncertainty in the potential parameters, and the resulting uncertainties incurred in using the potential to predict quantities of interest. A demonstration using the ensembles for three analytic potentials fitted for nickel to quantify uncertainties in elastic constants and thermal expansion coefficient is detailed in \sref{sec:application}. 

The uncertainty quantification component requires few external parameters to run. The number of potentials is specified, and the R value in (\ref{eq:rvalue}) is tuned. It is recommended that an initial investigation trialling a variety of R values is undertaken, with the objective being to accept approximately \SI{23}{\percent} of moves for optimal sampling of high dimensional cost landscapes \cite{roberts1997}. There is also the option for the user to alter the sampling temperature; $T = \alpha T_0$ where $\alpha$ is a scale factor. This alteration can be used to improve the convergence of the ensemble to the underlying distribution \cite{PhysRevLett.93.165501} and is demonstrated in \sref{sec:application} for the EAM potential. The calculation of the Hessian curvature, which relies upon a finite difference calculation of the cost space minima, also has a tunable perturbation parameter. The percentage perturbation to each parameter in the finite difference calculation can be tuned to ensure the curvature on the scale of the minimum basin is probed. In the case that a new minimum is found at any point in the process, the implementation outputs this new optimal parameter set and restarts the procedure. 

The ensemble generated is then used to calculate distributions of quantities of interest, through MD simulations using the ensemble members. The potential ensemble implementation can be used as part of an end-to-end potential fitting workflow, or can be instigated from a previously fitted \potfit potential as a stand-alone analysis. Full documentation for the implementation, as well as the code, is available on the \potfit website \footnote{\url{https://www.potfit.net}}. It should be noted that a systematic error in the reference data (estimated via methods described in e.g.\cite{Aldegunde2016}) is not accounted for. As discussed by Pernot and Cailliez\cite{Pernot17,Cailliez17}, the ensemble method assumes that the system uncertainty is dominated by model inadequacy rather than reference data uncertainty. As we show below, this is true for typical reference data and potential model combinations.

\section{Demonstration of uncertainty quantification for three analytic models }
\label{sec:application}	

\begin{figure}
	\centering
	\includegraphics[width=0.8\textwidth]{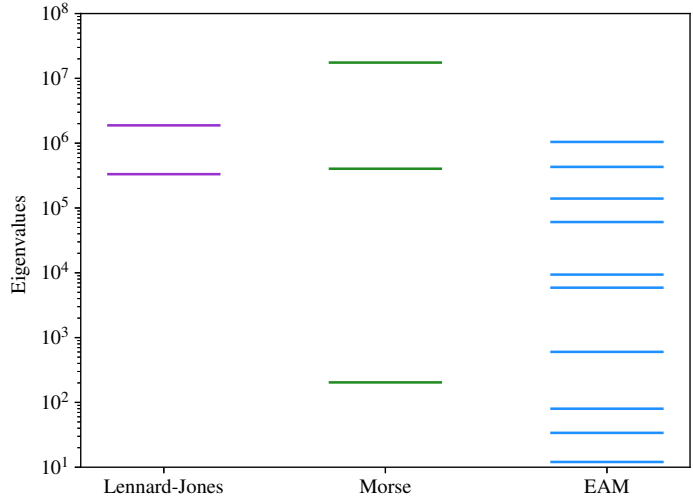}
	\caption{The spread in the eigenvalues of each Hessian for the potential models fitted to the reference set.}
	\label{fig:fig_1}
\end{figure}

We evaluate three analytic potential models of increasing complexity and compare their performance and uncertainty against DFT and experimental values. The potentials investigated were the Lennard-Jones (LJ), Morse and embedded atom method (EAM) potentials. The analytic functions and parameters fitted are detailed in \tref{table1}. Two pair potentials (LJ and Morse) were chosen as they contain two and three parameters respectively, thus the ensemble parameter output could be easily visualized in initial investigations. In the production of a potential for nickel, it is unlikely that a pair model would be chosen due to their simplistic nature and limited description of the system. They are applied in this exploration with the aim to highlight potential issues which may be encountered in practice, and to aid in building an understanding of the methods' scope. The \num{10}-parameter EAM potential is chosen, with the large parameter space (and encoded physics) expected to best reproduce the desired QoI with the least error. Furthermore, there are a variety of high quality EAM potentials for nickel in the literature \cite{doi:10.1080/14786430500333349,Mishin1999,Ackland1987,Angelo1995}. For the pair potential part of the EAM we have used a Morse function, for the embedding function, $F(n)$, the universal form proposed by Banerjea and Smith \cite{PhysRevB.37.6632} is used, and finally the transfer function, $\rho(r)$, is of the oscillatory form necessary for cubic metals as detailed in \cite{PhysRevB.53.14080}. A cut-off of \SI{10}{\angstrom} was used for each potential model, with the tails smoothed to converge to zero over \SI{0.75}{\angstrom}. The analytic functions for the potential models and smoothing function are detailed in \tref{table1}. The cut-off parameters impact the computational cost of the potential, which makes them indicators of the model complexity. Thus we decided to exclude them from the optimisation.

 \fulltable{
	\label{table1}The potential models and associated parameters. }
\br
Model& Analytic Function&Parameters \\
\mr
Lennard-Jones &  $V_{LJ}(r) = 4 \epsilon \left[\left(\frac{\sigma}{r}\right)^{12} - \left(\frac{\sigma}{r}\right)^6 \right] $& $\sigma,\,\epsilon$\\ \mr
Morse & $V_M(r) = D_e \Big[ \big( 1 - e^{-a(r-r_e)}\big)^2 -1\Big]$& $D_e,\, a,\, r_e$\\\mr
EAM & $E_i = \frac{1}{2} \sum\limits_{i\neq j}^N V_M(r_{ij}) + F(n_i)\quad \text{with} \quad n_i=\sum\limits_{j\neq i}^N\rho(r_{ij})$&$D_e,\, a,\, r_e$\\
\rule{0pt}{4ex}   & $\rho(r) = r^{-\beta}\,[1+a_1 \cos(\alpha r + \varphi)]$ & $a_1,\,\alpha,\,\varphi,\,\beta $\\
\rule{0pt}{4ex}   & $F(n) = F_0\, [1-\gamma \ln n]\, n^{\gamma}+F_1 n$& $F_0,\,\gamma,\,F_1$\\\mr
Smooth Cutoff &$ V_{SC}(r)=\Psi\left(\frac{r-r_c}{h}\right)V(r)\quad\qquad\text{where}\quad\Psi(x)=\frac{x^4}{1+x^4}$&$r_c=10,\,h=0.75$\\
\br
\endfulltable 

The potentials are fitted using \potfit to a nickel reference set comprising of \num{23} atomic configurations of \num{108}-atom fcc ($3 \times 3 \times 3$ unit cell) DFT and MD snapshots at a variety of temperatures and stresses. DFT simulations were performed in CASTEP\cite{CASTEP} using the PBE functional with a \SI{400}{\electronvolt} plane wave cut-off and a Monkhurst-Pack k-point grid of \SI{0.1}{\per\angstrom} spacing. This converges energies (forces) to within a tolerance of \SI{5e-5}{\electronvolt/atom} (\SI{0.05}{\electronvolt/\angstrom}) respectively. MD simulations were performed using LAMMPS \cite{PLIMPTON19951}. For each MD snapshot in the reference set, a DFT calculation was performed on the configuration to obtain the forces, energy and stresses used to fit the potentials. 
The ability of each fitted model to reproduce the reference data was observed to improve with increasing model complexity. The rms force errors for each fit were \SI{0.45}{\electronvolt/\angstrom} (LJ), \SI{0.16}{\electronvolt/\angstrom} (Morse) and \SI{0.11}{\electronvolt/\angstrom} (EAM), significantly higher than the DFT convergence levels. Similar trends were seen in the reproduction of reference energies and stresses.
 \Fref{fig:fig_1} shows the spread in the eigenvalues obtained for the three fitted analytic forms. With a difference of up to six orders of magnitude in the degree of curvature as denoted by the magnitudes of the eigenvalues, this illustrates the necessity for a sampling procedure which accounts for this variation in order to efficiently sample the underlying distribution. 

It is of note that the reference dataset was not particularly tailored to predict the elastic constants or thermal expansion coefficient. When fitting a potential for production use, the process typically involves multiple iterations of the reference dataset and potential fitting to tailor the regions of cost space explored for its intended use. The potentials generated are solely an illustration of the newly implemented potential ensemble method and are by no means suggested as new production-grade potentials to be used outside of this work in the modelling of nickel. 

 For each of the three fitted analytic potential forms, a \num{500}-member ensemble of potentials were obtained from MCMC samples output from the \potfit ensemble implementation. To ensure uncorrelated ensemble members, starting from the best fit potential, each sample was drawn after \num{50000} accepted steps with roughly \SI{23}{\percent} \cite{roberts1997}  of steps accepted for each analytic form through tuning of the value  of R in (\ref{eq:rvalue}). The decorrelation time was assessed from the autocorrelation of each parameter in an initial MCMC run, with a conservative decorrelation time of \num{50000} samples chosen. Sampling convergence was checked by ensuring reasonably smooth distributions in individual ensemble parameters and in all $2d-$projections of ensemble parameters.
 
 Sampling from the EAM potential landscape was performed at a reduced cost temperature of \SI{0.05}{T_0} in accordance with  \cite{PhysRevLett.93.165501}, due to the Markov chain leaving the minimum basin for higher temperatures. Outside, the Hessian calculated at the minimum is no longer valid, resulting in inefficient sampling. Reducing the temperature avoids these issues. The reduced sampling temperature for the EAM potential limits the insights from a direct comparison, but is not completely without merit; the scaled temperature results could be extrapolated to higher temperatures assuming a roughly quadratic basin in cost function space. We believe that foremost this behaviour is due to the selection of reference data the potential is fitted to. In the fitting of a potential for production use, the reference data is typically weighted and complemented with configurations generated using iterative improvements of the fitted potential, which might alleviate the issue. In case that does not provide a remedy, a different sampling strategy would need to be employed, e.g. Rieman Monte Carlo methods\cite{Girolami2011} or affine invariant samplers\cite{Goodman2010}. This might also allow incorporating prior information about the parameters into the ensemble generation process. 

We have chosen to investigate the performance of the potentials in reproducing the equilibrium lattice constant, the elastic constants $C_{11}$, $C_{12}$ and $C_{44}$, and the thermal expansion coefficient at \SI{300}{\kelvin}. Each analytic potential from each of the three ensembles was propagated through MD simulations to obtain the uncertainties for each potential model. 

The resulting uncertainties displayed in figures \ref{fig:fig_2} and \ref{fig:fig_3} are compared using box-whisker diagrams, illustrating the uncertainty in each quantity by the inter-quartile range (IQR). The box denotes the IQR, with whiskers extending $1.5\times$IQR beyond each quartile. The values obtained from the best fit (minimum cost) potential are shown in dashed purple. The notches indicate the confidence interval in the ensemble median (green) and the ensemble means are indicated as dotted red. The dotted black lines traversing the the entirety of each figure indicate the experimental values detailed in \Tref{table2}. Due to the presence of skewness and outliers, the uncertainties are reported using the resistant measures of sample median and IQR. Reporting uncertainties via the sample means and standard deviations will bias the reported quantities towards misleading ensemble outliers which tend to result from unlikely higher cost potential ensemble members.

 \fulltable{
	\label{table2}Comparison of results for QoI with their associated uncertainties (IQR). The equilibrium lattice constant is denoted $a$. $C_{11}$, $C_{12}$, $C_{44}$ are the elastic constants in Voigt notation, and $\alpha$ is the linear thermal expansion coefficient at \SI{300}{\kelvin}. The DFT value for $a$ is from a geometry optimized cell included in the reference data. }
\br
QoI&DFT/&\centre{2}{Lennard-Jones}&\centre{2}{Morse}&\centre{2}{EAM} \\
&experiment&median&IQR&median&IQR&median&IQR\\
\mr
$a$\,(\si{\angstrom}) & 3.51 & 3.14 &[2.94, 3.26]  &3.45&[3.33, 3.56] &3.52&[3.50, 3.54]  \\
$C_{11}\,(\si{\giga\pascal})$ \cite{Neighbours1952} &253  & 1503  & [1236, 2355] &224&[171, 315]&222&[180, 259]  \\ 
$C_{12}\,(\si{\giga\pascal})$ \cite{Neighbours1952}& 152  & 860 & [704, 1347] &158 &[123, 202]& 180&[130, 213] \\ 
$C_{44}\,(\si{\giga\pascal})$ \cite{Neighbours1952}  & 124 & 860 & [704, 1347] &158 &[123, 202]& 90&[81, 98] \\ 
$\alpha$\,(\si{\num{e-6}\, \kelvin^{-1}}) \cite{osti_4359714}& 14.4 &4.8&[3.9, 5.7]   & 8.2 &[7.4, 9.1]& 15.1  & [12.8, 17.6]\\ 
\br
\endfulltable

\subsection{Equilibrium lattice constant}

The lattice constants were found by minimising an fcc nickel lattice with a starting lattice parameter guess of \SI{3.5}{\angstrom} for each potential ensemble member. \Tref{table2} reports the uncertainties for each fitted analytic potential. 

The Lennard-Jones ensemble is unable to capture the correct lattice constant within the IQR. This is largely due to higher temperature configurations in the reference data to which the potential was initially fit. The Morse ensemble does manage to capture the correct value, but the larger spread in uncertainty for a simple \SI{0}{\kelvin} quantity again implies the potential is limited by the higher temperature configurations in the reference data. It is important to note that poor potential performance due to reference data selection is not always the culprit; if a model is known to have limitations in its ability to reproduce certain physical quantities due to its simplistic design, then any issues which arise may be down to an insufficient model choice. This is illustrated by the best-fit (minimal cost) LJ potential, where the first warning is in the large rms force error (\SI{0.45}{\electronvolt/\angstrom}). A second concern is the incorrect lattice constant prediction of \SI{3.23}{\angstrom}. This results in an ensemble of candidate potentials for an already insufficient fit, clearly indicating an insufficient model choice. Similarly, despite a more promising rms force error (\SI{0.16}{\electronvolt/\angstrom}), the Morse potential performs poorly in the prediction of lattice parameter, which also alludes to an insufficient model choice.

The EAM potential accounts for a non-linear dependency on the local environment through the embedding term. It appears that this is essential for a realistic prediction of the lattice constant given the set of reference data used: The EAM ensemble, albeit sampled with a reduced temperature, not only captures the DFT value within its error but also has a considerably constricted spread of the ensemble predictions compared to the pair potentials.

\subsection{Elastic constants}

\begin{figure}	
	\centering
	\includegraphics[width=0.8\textwidth]{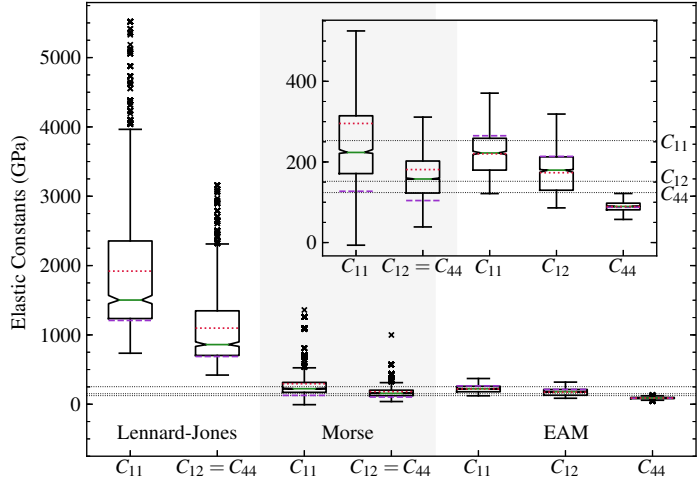}
	\caption{Elastic constants for each potential. The inset shows the Morse and EAM values without outliers for clarity. See text for details. }
	\label{fig:fig_2}
\end{figure}

The elastic constants are investigated to compare the restoring forces of the potentials to small perturbations of atomic positions. Calculation of the elastic constants for pair potentials is known to be unable to resolve the differences in $C_{12}$ and $C_{44}$ due to insufficient parameters to describe the off-diagonal tensor components. This is illustrated in \fref{fig:fig_2}, with only the EAM potential having distinct values for these elastic tensor components. It is noticeable that the best fit potential does not necessarily lie near the centre of the prediction interval. This highlights that there exist competing potentials of comparable suitability which may shift predictions in a particular direction away from the initially obtained best fit value.

The performance of the Lennard-Jones potential in the prediction of $C_{11}$ and $C_{12}=C_{44}$ is poor as expected. A very large spread in the predicted values for both, all of which fail to reproduce the experimental values is expected of an ill-fit two parameter potential. Due to the outliers in the fit we observe a stark difference between the mean and median values. When dealing with a model which clearly fails to correctly reproduce the elastic constants, this tells us little more than there is a disagreement in predictions from ensemble members. Further to this, the best fit predictions are also vastly incorrect. Together these observations clearly demonstrate the known limitations of such a simple potential, illustrating results in line with an insufficient choice of model. 

The Morse potential is able to capture the expected experimental values for both $C_{11}$ and $C_{44}$, although still unable to resolve the $C_{12}$, $C_{44}$ difference due to the pair potential nature. The spread in the ensemble mean and median predictions illustrate a disagreement in predictions from candidate potentials, which again alludes to insufficient model choice. This is another example of poor model selection despite promising initial predictions of the diagonal elasticity elements. 

The EAM potential is able to capture the expected trend for the three constants but does not capture the expected value for $C_{44}$ in the uncertainty. The initial failure of the fitted EAM potential to achieve the expected off-diagonal $C_{44}$ component could be rectified through improvements to the reference data. Comparison of the best-fit tensor component predictions with the ensemble mean and median illustrate the performance of suitable alternative candidate potentials. \Fref{fig:fig_2} focuses on demonstrating the output of the \potfit uncertainty quantification, for a comparison of production grade EAM potentials in predicting the elastic constants the reader is referred to \cite{EAM_pot_comparison}.

\subsection{Thermal expansion coefficient}

\begin{figure}
	\centering
	\includegraphics[width=0.8\textwidth]{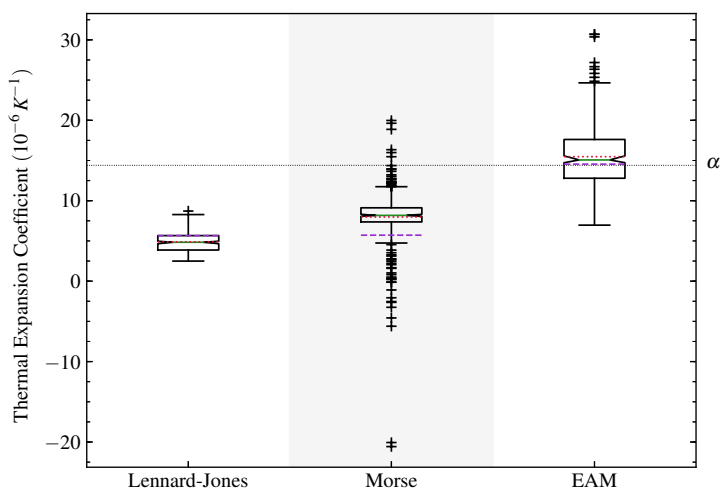}
	\caption{The thermal expansion coefficient at \SI{300}{\kelvin} for each potential. See text for details. }
	\label{fig:fig_3}
\end{figure}

Finally the thermal expansion coefficient for nickel at \SI{300}{\kelvin} was calculated. The linear expansion of the solid is investigated to compare the energetic contributions to the system provided by the potentials. The thermal expansion coefficient was calculated by evaluating the length change of crystalline nickel for five temperatures, symmetrically distributed around \SI{300}{\kelvin} at \SI{20}{\kelvin} intervals. A curve is then fitted to the results using regression to find the thermal expansion coefficient at \SI{300}{\kelvin}.

On first inspection the results in \fref{fig:fig_3} may misleadingly imply that the pair potentials outperform the EAM, due to the small spread in the ensemble members. However upon closer inspection, the predicted pair values in-fact fail to capture the correct value within the IQR and even within the tails of the distribution. The significant outliers in the Morse predictions again imply a disagreement in predictions of ensemble members. The EAM potential does manage to bound the correct value within the uncertainty, although a large spread in the uncertainty is again likely a result of the choice of reference data to which it has been fit. 

The uncertainties for the thermal expansion coefficient demonstrate the importance of looking at the predictions of a selection of relevant QoI when evaluating the suitability of a potential model. In \tref{table2} the predicted lattice constant and uncertainty bounds for the Lennard-Jones potential fail to capture the simple equilibrium quantity despite a large uncertainty. This unsuccessful prediction, combined with the high relative uncertainty in the thermal expansion coefficient, illustrate an example of poor model selection. This leads to a caveat in the application of the ensemble framework: its application should not be used as a means to bypass an informed fitting of potentials. In the case of the Lennard-Jones potential, the initial best fit performed poorly in predicting the correct forces in the reference set. Furthermore, the incorrect prediction of the lattice parameter by the best fit potential, and the resulting ensemble, imply that the model is insufficient. Failure to assess the model suitability at each stage of the fitting process can result in misleading uncertainties, as we have attempted to illustrate in this case.

\section{Conclusion}
 
The \potfit potential fitting workflow has been enhanced to generate an ensemble of potentials encapsulating the uncertainty of the correct parameters. This allows a propagation of this uncertainty to quantities of interest of molecular dynamics simulations. The ensemble is generated by sampling the cost landscape using sampling techniques developed for sloppy models. In this work we demonstrated a preliminary uncertainty quantification of quantities of interest for three distinct effective potential models. The implementation enables users to quantify the uncertainty of simulation values incurred by the choice of potential parameters. In future, the current implementation may be improved by using more efficient sampling algorithms.

 The ensemble method can be used to build an understanding of the impact of parameter uncertainty on the precision of quantities of interest. As our results illustrate, users must be aware that this method provides a lower bound of the error bars; some quantities might not be described well by a potential model. This is why it is mandatory that users of the ensemble method implementation diligently evaluate the suitability of the model throughout the fitting (as reported by rms errors in reproducing reference data) and uncertainty quantification process.
 
This is a further puzzle piece towards reproducible and transparent MD simulations -- an effective potential should not exist on its own, but rather together with its implementation (as e.g.\ provided by the OpenKIM framework \cite{0965-0393-25-1-014001}), its reference data and its uncertainties (\potfit + UQ). This integration is also a step toward predictive simulations, i.e.\ with error bounds determined \emph{a priori}.

A potential future line of enquiry opened by this work is to investigate how the ensemble information may be used efficiently in production simulations. While ensemble simulations are trivially parallel and scale perfectly with the number of ensemble members, reducing the number of simulations performed may still be desirable, e.g.\ by intelligently selecting and weighting potentials in the ensemble. Similarly the determined uncertainties could inform the choice of reference data. Closing this feedback loop may lead to further improved classical effective potentials and trustworthy simulation results with quantified uncertainties.

\ack

The authors gratefully acknowledge support from the Engineering and Physical Sciences Research Council (EPSRC) Grant No.\ EP/M508184/1.
Simulations were performed using the high performance computing facilities from the Warwick Scientific Computing Research Technology Platform (SC RTP). We would like to thank Daniel Schopf for support with \potfit, and Mingjian Wen for discussions on the nature of the EAM potential cost landscape. 

The \potfit ensemble method implementation is available at \\ \url{https://github.com/potfit/potfit/tree/ensemble_method_uq}.  All data necessary to reproduce this research can be accessed on the Warwick Research Archive Portal (WRAP) at \url{http://wrap.warwick.ac.uk/114795}, and used under the Creative Commons Attribution licence. 

\section*{References}

\bibliographystyle{iop}
\bibliography{SL_Paper_arXiv.bib}

\end{document}